\documentclass[showpacs,showkeys,preprintnumbers,amsmath,amssymb]{revtex4}

\usepackage{graphicx}

\usepackage{dcolumn}
\usepackage{bm}

\providecommand{\beqa}{\begin{eqnarray}}
 \providecommand{\bf}{\mathbf}
 \providecommand{\rm}{\mathrm}
\providecommand{\eeqa}{\end{eqnarray}}

 \def\A{{\alpha_k}}
 \def\B{{\beta_k}}
\def\OA{{\overline{\alpha_k}}}
\def\OB{{\overline{\beta_k}}}

\def\bk{{\bf k}}

\def\Mpl{M_{pl}}

\def\mE{{\mathcal{E}}}
\def\tk{{\tilde{k}}}
\begin{document}

\preprint{\rightline{MAD-TH-10-08}}
\preprint{\rightline{UUITP-42/10}}
\title{A Note on Calm Excited States of Inflation}

\author{Amjad Ashoorioon}
\email{amjad.ashoorioon@fysast.uu.se}

\affiliation{Institutionen f\"{o}r fysik och astronomi
Uppsala Universitet, Box 803, SE-751 08 Uppsala, Sweden}
\author{Gary Shiu}
\email{shiu@physics.wisc.edu}
\affiliation{Department of Physics, University of Wisconsin,
Madison, WI 53706, USA}


\date{\today}

\begin{abstract}

We identify a two-parameter family of excited states within slow-roll inflation for which either the corrections to the two-point function or the characteristic signatures of excited states in the three-point function -- {\it i.e.} the enhancement for the flattened momenta configurations-- are absent. These excited states may nonetheless violate the adiabaticity condition maximally. We dub these initial states of inflation  {\it calm excited states}. We show that these two sets do not intersect, {\it i.e.}, those that leave the power-spectrum invariant can be distinguished from their bispectra, and vice versa. The same set of calm excited states that leave the two-point function invariant for slow-roll inflation, do the same task for DBI inflation. However, at the level of three-point function, the calm excited states whose flattened configuration signature is absent for slow-roll inflation, will lead to an enhancement for DBI inflation generally, although the signature is smaller than what suggested by earlier analysis.
This example also illustrates that imposing the Wronskian condition is important for obtaining
a correct estimate of the non-Gaussian signatures.

\end{abstract}


\maketitle
\section{Introduction}

Primordial non-Gaussianity is among the most promising probe of our early universe \cite{Komatsu:2009kd}.
Although current cosmological data is consistent with the primordial density fluctuations being Gaussian,
any observed departure from a Gaussian spectrum
will likely point us to some
interesting microphysics.
In the simplest vanilla version of inflation, namely, slow-roll inflation with a Bunch-Davies intial state, the primordial non-Gaussianity produced was shown to be suppressed by  the slow-roll parameters \cite{Maldacena:2002vr,Acquaviva:2002ud} and is thus unobservably small.
Nonetheless, well motivated deviation from this standard picture can lead to a detectable signal in current and upcoming experiments.
For example, even within the context of single-field inflation,
 non-canonical kinetic terms
can generate large bispectra (3-point function) of the equilateral shapes \cite{Chen:2006nt,Cheung:2007st}; a non Bunch-Davies initial state of inflation can boost the folded/flattened momentum configuration \cite{Chen:2006nt,Holman:2007na,Meerburg:2009ys,Meerburg:2009fi}; and features in the Lagrangian can result in a large bispectrum with oscillatory running \cite{Chen:2006xjb,Chen:2008wn}. A combination of the above effects realized as resonant folded non-Gaussianity was considered in \cite{Chen:2010bka}. Multifield inflationary models open up even further possibilities, such as curvatons \cite{Lyth:2002my},
turning \cite{Vernizzi:2006ve,Huang:2007hh,Langlois:2008qf,Arroja:2008yy,Chen:2008ada,Byrnes:2008zy,Chen:2010qz,Ashoorioon:2008qr}
or bifurcating \cite{Naruko:2008sq,Li:2009sp} trajectories, thermal effects \cite{Moss:2007cv,Chen:2007gd} and etc \footnote{For a recent review on non-Gaussian effects for different inflationary models, see \cite{Chen:2010xka} which also contains an organized classification of different types of non-Gaussianity.}.

Among the various possibilities, the choice of initial state of inflation is perhaps
the least understood from a microphysical point of view.
While the other scenarios mentioned above can be analyzed
within well defined field theoretical models, the choice of initial conditions for inflation require a
detailed understanding of short distance physics.
The fact that the universe may have inflated much more than needed to solve the flatness and horizon problems of standard cosmology implies that the CMB scales today may have been
not far from the Planck scale at
 the beginning of inflation. This is sometimes known as the ``trans-Planckian problem" of inflationary cosmology \cite{Brandenberger:1999sw}, though we hasten to stress here that
 this problem is not specific to Planck scale physics.
 We will hereforth refer to the Trans-Planckian problem as the question of how to formulate inflation when there exist a short distance scale $M$ (e.g., the string scale $M_s$) not far from the Hubble scale of inflation.
 Our understanding of physics at such high energies is rather limited at present.
One may however view this  UV problem of inflation as an opportunity, as CMB and other cosmological measurements can provide us a way to probe high scale physics.
The short distance modifications of inflation can affect the
dynamics of
 the primordial perturbations \cite{Martin:2000xs,Kempf:2000ac}  as effective field theory arguments would suggest \cite{Kaloper:2002uj,Shiu:2002kg},
 or their initial conditions \cite{Danielsson:2002kx}. While a UV completion of inflation that allows us to unambiguously determine its initial state is not yet available, one often resorts to a phenomenological approach.
In \cite{Danielsson:2002kx}, a class of initial states
 was motivated based on the notion of vacuum ({\it i.e.}, minimizing particle production) in a theory with a UV cutoff.
The computed power spectrum turns out to be different from that obtained with the Bunch-Davies vacuum. Subsequent work in formulating such initial state effects in terms of boundary effective field theory \cite{BEFT} further exemplified this fact.
Thus, it is generally expected to see modification of the power spectrum with any non Bunch-Davies initial state \cite{Chen:2006nt,Kempf:2000ac},
though the deviation may not always be
large enough to be observed. Having an excited state as the initial condition for inflation can be justified
as a
 pre-inflationary phase could have excited the inflaton fields from their vacuum \cite{Powell:2006yg,Sarangi:2006yy} or some non-adiabatic process, at the time when the energy scale of inflation was around $M$, could excite the vacuum to an excited state \cite{Holman:2007na,Ashoorioon:2006wc,Ashoorioon:2008qr}. The effect of such excited states are usually discussed in terms of modulated oscillations on the power spectrum. However, as originally pointed out in \cite{Chen:2006nt}, the effect of such excited states may be more conspicuous at the level of three-point function probes, especially for slow-roll inflationary models with higher derivative correction terms  \cite{Chen:2006nt,Holman:2007na,Meerburg:2009ys} or in the context of DBI inflation \cite{Chen:2006nt,Meerburg:2009fi}.

In this note, we address a related but different question: can such initial state effects be hidden? We identify excited states that can deviate from the Bunch-Davies vacuum as much as it is allowed by the constraints from backreaction, but do not leave {\it any} signature at the level of the two point function. We call these excited states ``calm excited states". We show that all these excited states are revealed at the level of the three-point function through their enhanced
contribution in the flattened momentum configurations.
Our analysis also suggests that an approximation which was
made
 in previous studies \cite{Holman:2007na,Meerburg:2009ys,Meerburg:2009fi}, in which the first Bogolyubov coefficient was set to its Bunch-Davies value,
may lead to an inaccurate estimate of such enhancement.
 We argue the same set of calm excited states would leave the power-spectrum invariant in DBI inflation. For slow-roll inflation, one can also identify excited states that again maximally violate the adiabaticity condition but the enhancement for flattened configurations is {\it identically} zero. However, in DBI inflation, it it is not possible for an excited state to give  vanishing enhancement to the flattened configuration.

The outline of this short note is as follows: in the first part we construct initial excited states that leave the two-point function of slow-roll inflation intact. We then compute the three-point function for a general excited state without resorting to any approximation for the Bogolyubov coefficients by imposing properly the Wronskian condition.  We further identify the calm excited states in DBI inflation. Finally we show that in DBI inflation any excited state that maximally violates the adiabaticity condition could be {\it observed} by non-gaussianity probes unless the scale of new physics, $M$, is higher than $10^{-2}M_{pl}$. Explicit expressions for the bispectrum with general excited state in DBI inflation are relegated to the appendix. We end with a brief conclusion.

\section{Slow-roll Inflation
}
\subsection{Calm Excited States with no Modification to the Power Spectrum}

\subsubsection{Power Spectrum}

We would like to know if one can modify the initial conditions for perturbations that correspond to excited states, but still leave the power spectrum intact. Let us assume for simplicity that the inflationary background is de-Sitter space, $a(\eta)=-\frac{1}{H\eta}$. The perturbations in such a background satisfy the following differential equation
\begin{equation}\label{u-eq}
u''_k+(k^2-\frac{2}{\eta^2})u_k=0.
\end{equation}
The full solution to this differential equation is
\begin{equation}\label{u-sol}
u_k(\eta)=\A~(-\eta)^{1/2}  H_{3/2}^{(1)}(-k\eta)+\B~(-\eta)^{1/2} H_{3/2}^{(2)}(-k\eta).
\end{equation}
The term proportional to $H_{3/2}^{(1)}(-k\eta)$ behaves like positive-frequency wave in the infinite past and the other term behaves like the negative frequency one. The Wronskian condition, $u^{\ast} u'-u u'^{\ast}=-i$, will result in the following relation between $\A$ and $\B$
\begin{equation}\label{Wronskian}
|\A|^2-|\B|^2=\frac{\pi}{4}.
\end{equation}
The power spectrum for the general solution \eqref{u-sol} is
\begin{equation}\label{power-spectrum}
P_S=\frac{|\A-\B|^2 H^2}{\pi^3\epsilon}
\end{equation}
We would like to know if it is possible to obtain an unmodified scale-invariant power spectrum, $P_S=\frac{H^2}{4\pi^2 \epsilon}$ starting with non-zero $\A$ and $\B$, {\it i.e.} if it is possible that
\begin{equation}\label{AB-scale-invariant}
|\A-\B|^2=\frac{\pi}{4}.
\end{equation}
$\A$ and $\B$ are in general complex variables. We write them as
\begin{eqnarray}
  \A &=& x_1+i x_2, \\
  \B &=& y_1 +i y_2.
\end{eqnarray}
Eqs. \eqref{AB-scale-invariant} and \eqref{Wronskian} could then be written as
\begin{eqnarray}\label{relation-x-y}
 \nonumber
  x_1^2+y_1^2-2 x_1 y_1+x_2^2+y_2^2-2 x_2 y_2 &=& \frac{\pi}{4}, \\
  x_1^2+x_2^2 -y_1^2-y_2^2 &=&\frac{\pi}{4}.
\end{eqnarray}
From the above two equations one can solve for $x_1$
\begin{equation}\label{x1}
x_1=\frac{y_1^2+y_2^2}{y_1}-\frac{x_2 y_2}{y_1}.
\end{equation}
Plugging this back to the second equation of \eqref{relation-x-y} and solving for $x_2$, one obtains  a second order equation for $x_2$ whose solutions are:
\begin{equation}\label{x2-first}
x_2^{(1)}=y_2+\frac{|y_1|\sqrt{\pi}}{2\sqrt{y_1^2+y_2^2}},
\end{equation}
\begin{equation}\label{x2-second}
x_2^{(2)}=y_2-\frac{|y_1|\sqrt{\pi}}{2\sqrt{y_1^2+y_2^2}}.
\end{equation}
One can then solve for $x_1$ using \eqref{x2-first} and \eqref{x2-second} and respectively
and find two solutions for $x_1$:
\begin{equation}\label{x1-first}
x_1^{(1)}=y_1-y_2~{\rm sign}(y_1)\frac{\sqrt{\pi}}{2\sqrt{y_1^2+y_2^2}},
\end{equation}
\begin{equation}\label{x1-second}
x_1^{(2)}=y_1+y_2~{\rm sign}(y_1)\frac{\sqrt{\pi}}{2\sqrt{y_1^2+y_2^2}}.
\end{equation}
Let us  mention here some illustrative examples of calm excited states: $\B$ can be chosen such that the number of particle that are produced during inflation do not backreact and spoil inflation. In principle, the real and imaginary parts of $\B$ do not have to be related, but for simplicity,
let's
consider a first example where they do:
\begin{equation}\label{B}
\B=\sigma (1+i).
\end{equation}
From the bounds on backreaction, $\sigma$ is expected to be smaller than \cite{Holman:2007na}
\begin{equation}\label{sigma-bound}
\sigma \lesssim {\rm min}\{\sqrt{\epsilon}\frac{H M_{\rm P}}{M^2},\sqrt{\epsilon \eta'}\frac{H M_{\rm P}}{M^2}\}.
\end{equation}
$M$ is the momentum cut-off, below which the effective field theory description is valid. The bound on $\B$ comes from the fact that the energy density of the nearly massless quanta of the inflaton has to be less than $M_{\rm p}^2 H^2$. Also the produced energy should not violate the slow-roll condition, hence the appearance of the first and second slow-roll parameters, $\epsilon$ and $\eta'$. These two considerations will lead to the above bound on $\sigma$. In obtaining the results, one has to assume that $\B\rightarrow 0$ for $k>M(a(\eta_0))$, where $\eta_0$ is the conformal time the physical momentum of the comoving mode becomes equal to the cutoff of the effective field theory, {\it i.e.} $k/a(\eta_0)\simeq M$ \cite{Holman:2007na}\cite{Boyanovsky:2006qi}. From the above consideration, one can see that $\sigma$ is not necessarily very small. For example, for an inflationary potential with $V=\frac{1}{2}m^2 \phi^2$, with $H \simeq 3.6 \times 10^{-5}$ and $\epsilon\simeq \eta'\simeq 0.01$, for $M=10 H$, $\sigma\simeq 1.36$ in the above example. Note that earlier analysis in the literature which
constraint $M$ to be greater than $100 H$ \cite{Okamoto:2003wk,Easther:2004vq} do not apply to the the above example
as
the calm excited states
are special in that they
 do not leave any signature in the power spectrum.

For the above choice of $\B$, eq.\eqref{B}, the first Bogolyubov coefficient could be obtained from eqs. \eqref{x2-first}-- \eqref{x1-second} to have a calm excited state:
\begin{equation}\label{A1}
\A^{(1)}=\left(\sigma-\frac{\sqrt{\pi}}{2\sqrt{2}}\right)+i \left(\sigma+\frac{\sqrt{\pi}}{2\sqrt{2}} \right)
\end{equation}
and
\begin{equation}\label{A2}
\A^{(2)}=\left(\sigma+\frac{\sqrt{\pi}}{2\sqrt{2}}\right)+i \left(\sigma-\frac{\sqrt{\pi}}{2\sqrt{2}} \right).
\end{equation}
Starting with either pair of $(\A^{(1)},\B)$ and $(\A^{(2)},\B)$, one
obtains
the same power spectrum
as
that for the Bunch-Davies initial state.

Another example which maximally deviates from the Bunch-Davies vacuum but still satisfies the bound from the backreaction is given by:
\begin{equation}\label{B2}
\B=\sigma. \end{equation}
The Wronskian condition then implies the following two solutions for $\A$:
\begin{equation}\label{A21}
\A^{(1)}=\sigma+i\frac{\sqrt{\pi}}{2},
\end{equation}
and
\begin{equation}\label{A22}
\A^{(2)}=\sigma-i\frac{\sqrt{\pi}}{2}.  \end{equation}
The fact that for this case the imaginary part of $\beta_k$ is zero will be of interest in the next section. As we will see, following \cite{Holman:2007na}, one would obtain zero enhancement for flattened configurations for the above excited state, whereas the actual enhancement is nonzero if
the $\A$ coefficient
satisfies
 the Wronskian condition \eqref{Wronskian} properly.

\subsubsection{Three-point function}

Let us calculate the three-point function for the above excited states to see if they modify the three-point function, and if so, how large the effect will be. One can calculate the Wightman function for the solution \eqref{u-sol}
 \begin{equation}\label{Wightman}
 G_{k}^{>}(\eta,\eta')\equiv\frac{H^2}{{\dot{\phi}}^2} \frac{u_k(\eta)}{a(\eta)}\frac{u_k^{\ast}(\eta')}{a(\eta')}
 \end{equation}
 The three-point function could be derived from the Wightman function through the following integral \cite{Maldacena:2002vr}:
 \begin{eqnarray}\label{zeta3}
\langle \zeta_{\overrightarrow{k_1}}\zeta_{\overrightarrow{k_2}}\zeta_{\overrightarrow{k_3}}\rangle&=& -i (2\pi)^3 \delta^3\left(\sum \overrightarrow{k_i} \right) \left(\frac{\dot{\phi}}{H}\right)^4 M_P^{-2} H \int_{\eta_{0}}^0 d\eta \frac{1}{k_3^2} (a(\eta) \partial_\eta G_{k_1}^{>}(0,\eta)) (a(\eta) \partial_\eta G_{k_2}^{>}(0,\eta)) (a(\eta) \partial_\eta G_{k_3}^{>}(0,\eta))\nonumber \\&+&{\rm permutations}+{\rm c.c.}
 \end{eqnarray}
To compute the three-point function, the first argument of the Wightman function has to be set to zero, which corresponds to the moment the mode is outside the horizon, and then differentiate it with respect to the second argument. We obtain
\begin{equation}\label{aGprime}
a(\eta) \partial_{\eta}G_{k}^{>}(0,\eta))=\frac{H^3}{\dot{\phi}^2}\frac{2(\A-\B)(-\overline{\A}\exp(i k\eta)+\overline{\B}\exp(-ik\eta))}{\pi k}
\end{equation}
Plugging this result back to equation \eqref{zeta3}, one obtains:
\begin{eqnarray}\label{zeta4}
\langle \zeta_{\overrightarrow{k_1}}\zeta_{\overrightarrow{k_2}}\zeta_{\overrightarrow{k_3}}\rangle&=& \delta^3\left(\sum \overrightarrow{k_i} \right) \frac{64 H^6 (\A-\B)^3}{\dot{\phi}^2 k_1 k_2 k_3^3 M_P^2}\left(\frac{\overline{\A}^3}{k_t}\exp(i k_t \eta)-\OA^2\OB \sum_{j=1}^{3}\frac{\exp(i \tilde{k}_j \eta)}{\tilde{k}_j}-\OA\OB^2 \sum_{j=1}^{j=3} \exp(-i \tilde{k}_j) \right.\nonumber \\
&+&\OB^3 \frac{\exp(-i k_t \eta)}{k_t}\Bigg)\Bigg|_{\eta=\eta_0}^{0}+{\rm c.c.}+{\rm permutations}
 \end{eqnarray}
 where $\eta_0$ is the moment at which the physical momentum becomes equal to the physical cutoff, {\it i.e.} $\eta_0=\frac{M}{H k}$. $k_t$ is the sum of the magnitudes of the momenta $k_t=k_1+k_2+k_3$ and $\tilde{k}_j=k_t-2k_j$. Upon
  integration, one obtains:
 \begin{equation}\label{zeta5}
 \langle \zeta_{\overrightarrow{k_1}}\zeta_{\overrightarrow{k_2}}\zeta_{\overrightarrow{k_3}}\rangle=\frac{64 H^6}{\dot{\phi}^2 k_1 k_2 k_3}\left[\frac{(1-\cos(k_t \eta_0))}{k_t}C_1+i\frac{\sin{k_t}\eta_0}{k_t}C_2+C_3\sum_{j=1}^{3}\frac{(1-\cos(\tilde{k}_j)\eta_0)}{\tilde{k}_j}+C_4\sum_{j=1}^{3} i\frac{\sin{\tilde{k}_j\eta_0}}{\tilde{k}_j}\right](\frac{1}{k_1^2}+\frac{1}{k_2^2}+\frac{1}{k_3^2})
 \end{equation}
 where
 \begin{eqnarray}
 C_1 &=& (\A-\B)^3 (\OA^3+\OB^3)+(\OA-\OB)^3 (\A^3+\B^3) \\
 C_2&=& (\OA-\OB)^3 (\A^3-\B^3)+(\A-\B)^3 (\OB^3-\OA^3) \\
 C_3 &=& (\A-\B)^3 (-\OA\OB^2-\OA^2\OB)+(\OA-\OB)^3 (-\A\B^2-\A^2\B)) \\
 C_4 &=& (\A-\B)^3 (\OA^2\OB-\OA\OB^2)+(\OA-\OB)^3 (\A\B^2-\A^2 \B)
 \end{eqnarray}
As it was noticed originally in \cite{Chen:2006nt}, the fact that $\tilde{k}_j$  in the denominator of \eqref{zeta5} vanishes, enhances the contribution of these flattened configurations with respect to the Bunch-Davies vacuum. The divergence in the $\tilde{k}_j\rightarrow 0$ is removed by considering that the cut-off is only relevant for $\tilde{k}_j\simeq \frac{1}{\eta_0}$. Thus for the configurations in which one of the $\tilde{k}_j$ vanishes, one obtains:
\begin{equation}\label{zeta6}
 \langle \zeta_{\overrightarrow{k_1}}\zeta_{\overrightarrow{k_2}}\zeta_{\overrightarrow{k_3}}\rangle \simeq \frac{64 H^6}{\dot{\phi}^2 k_1 k_2 k_3} \left[\frac{(1-\cos(k_t \eta_0))}{k_t}C_1+i\frac{\sin{k_t}\eta_0}{k_t}C_2+i C_4 \eta_0 \right] (\frac{1}{k_1^2}+\frac{1}{k_2^2}+\frac{1}{k_3^2})
\end{equation}
The first two terms will change the normalization and shape dependence of the three-point function. The last term is the enhancement which is important for the flattened configurations. The relative enhancement factor with respect to the standard result is:
\begin{equation}\label{rel-enhance}
\left.\frac{\Delta  \langle \zeta_{\overrightarrow{k_1}}\zeta_{\overrightarrow{k_2}}\zeta_{\overrightarrow{k_3}}\rangle}{ \langle \zeta_{\overrightarrow{k_1}}\zeta_{\overrightarrow{k_2}}\zeta_{\overrightarrow{k_3}}\rangle}\right|_{\tilde{k}_j=0}\approx i C_4 k_t \eta_0= i C_4\frac{k_t}{a(\eta_0)H}
\end{equation}
The result is enhanced by the
ratio of
the
physical momentum to the Hubble scale at the beginning of inflation. The factor $C_4$ is what has been already approximated with $\Im(\B)$ in \cite{Holman:2007na},
by
setting $\A= \frac{\sqrt{\pi}}{2}$ \footnote{Note the in \cite{Holman:2007na} the solutions to the mode equation are identified such that $\A=1$ corresponds to the solution that approaches the Bunch-Davies vacuum as $k\eta\rightarrow -\infty$. This is different than the the convention of \cite{Chen:2006nt} by an $i$ factor.}. Here, we do not resort to this approximation. $C_4$ could be factorized as
\begin{equation}\label{C4-2}
C_4=(|\A|^2 -|\B|^2)(|\A-\B|^2)(\A\OB-\OA\B)
\end{equation}
Using the Wronskian condition \eqref{Wronskian} and the condition \eqref{AB-scale-invariant}, we can rewrite it as
\begin{eqnarray}\label{}
C_4&=&\frac{i\pi^2}{8} \Im(\A\OB)\\
&=&\frac{i\pi^2}{8} (x_2 y_1-x_1 y_2)
\end{eqnarray}
Note that setting $\A=\frac{\sqrt{\pi}}{2}$, we recover that $C_4$ is proportional to $\Im(\B)$, as stated in \cite{Holman:2007na}. Using the pair of expressions (\eqref{x2-first},\eqref{x1-first}) or (\eqref{x2-second}, \eqref{x1-second}), one respectively finds:
\begin{eqnarray}
  C_4 &=& \pm \frac{i\pi^2}{16} \left(\frac{y_1 |y_1|\sqrt{\pi}}{\sqrt{y_1^2+y_2^2}}+ \frac{y_2^2~{\rm sign}(y_1)\sqrt{\pi}}{\sqrt{y_1^2+y_2^2}}\right) \nonumber \\
      &=& \pm \frac{i\pi^{\frac{5}{2}}}{16} \sqrt{y_1^2+y_2^2}~{\rm sign}(y_1)
\end{eqnarray}
From the last expression for $C_4$ one can see that  in contrast to \cite{Holman:2007na},
the enhancement for the flattened configuration is never zero for the calm excited states, even when the second Bogolyubov coefficient is completely real.
Therefore the flattened configurations are enhanced even in the case of  \eqref{B2}.
The enhancement is proportional to
\begin{equation}\label{B2-enhancemnt}
\left.\frac{\Delta  \langle \zeta_{\overrightarrow{k_1}}\zeta_{\overrightarrow{k_2}}\zeta_{\overrightarrow{k_3}}\rangle}{ \langle \zeta_{\overrightarrow{k_1}}\zeta_{\overrightarrow{k_2}}\zeta_{\overrightarrow{k_3}}\rangle}\right|_{\tilde{k}_j=0}\approx \mp \frac{1}{16} \sigma \pi^{\frac{5}{2}} k_t\eta_0
\end{equation}
This illustrates the importance of the Wronskian condition
in getting the correct estimation of non-Gaussianities.


\subsection{Excited States with No Enhancement for the Flattened Configurations in Slow-roll Inflation}

In the previous sections, we investigated whether there exist calm excited states that leave the two-point function unmodified and noticed that all these excited states will lead to an enhancement of the flattened configuration. Now that we are equipped with the general expression for the enhancement factor, let us see if we can also identify a family of  excited states that do not lead to {\it any} enhancement of the bi-spectrum for the flattened configurations. Such excited states still satisfy the Wronskian condition, eq. \eqref{Wronskian}. However instead of equation  \eqref{AB-scale-invariant}, which was necessary for obtaining an intact two-point function, we would like to assume that the enhancement for the flattened configurations is absent, namely we have
\begin{equation}\label{C4-zero}
C_4=(|\A|^2 -|\B|^2)(|\A-\B|^2)(\A\OB-\OA\B)=0
\end{equation}
which can be simplified further using the Wronskian condition and writing the expressions in terms of real and imaginary parts of $\A$ and $\B$
\begin{equation}\label{C4-zer-rewritten}
C_4=\frac{\pi}{4}(|\A-\B|^2)(\A\OB-\OA\B)=-i \frac{\pi}{2}(y_2 x_1-x_2 y_1)(y_2^2+x_1^2-2 x_2 y_2+x_2^2-2 y_1 x_1+y_1^2).
\end{equation}
Finding such nontrivial excited states is therefore contingent upon whether non-trivial solutions for the following pair of equations exist:
\begin{eqnarray}
  x_1^2+x_2^2 -y_1^2-y_2^2 &=&\frac{\pi}{4}\label{eq1-Wronskian}\\
 (y_2 x_1-x_2 y_1)(y_2^2+x_1^2-2 x_2 y_2+x_2^2-2 y_1 x_1+y_1^2)  &=& 0 \label{eq2-c4-equal-to-zero}
\end{eqnarray}
The equation \eqref{eq2-c4-equal-to-zero}, by itself constitutes of two equations:
\begin{eqnarray}\label{c4zero}
y_2 x_1-x_2 y_1=0 \label{c4-zeo-first-eq}\\
y_2^2+x_1^2-2 x_2 y_2+x_2^2-2 y_1 x_1+y_1^2=0 \label{c4-zeo-second-eq}
\end{eqnarray}
Solving \eqref{eq1-Wronskian} and  \eqref{c4-zeo-second-eq}, one would obtain solutions for $x_1$ and $x_2$ (in terms of $y_1$ and $y_2$) that are not real. Therefore, these pairs of equations would not lead to any meaningful results. Solving the other pair of equations \eqref{eq1-Wronskian} and \eqref{c4-zeo-first-eq} together yield the following results for $x_1$ and $x_2$
\begin{eqnarray}
  x_2 &=& y_2\left(\frac{\frac{\pi}{4}+y_1^2+y_2^2}{y_1^2+y_2^2}\right)^{1/2} \\
  x_1 &=& y_1\left(\frac{\frac{\pi}{4}+y_1^2+y_2^2}{y_1^2+y_2^2}\right)^{1/2}
\end{eqnarray}
These two equations determine a two-parameter family of solutions for which the enhancement from flattened configurations are absent.
Two example of such states are
\begin{eqnarray}
\B &=& \sigma (1+i)  \\
\A &=& (\frac{\pi}{8}+\sigma^2)^{1/2}+i (\frac{\pi}{8}+\sigma^2)^{1/2}
\end{eqnarray}
and
\begin{eqnarray}\label{c4-zero-ex2}
  \B &=& \sigma \\
  \A &=& \left(\frac{\pi}{4}+\sigma^2\right)^{1/2}
\end{eqnarray}

One can work out the power spectrum for this class of calm excited states and see that
the power spectrum takes the form
\begin{equation}\label{pwr-calm-excited}
P_S=\frac{H^2}{(2 \pi)^2\epsilon}{\left(\frac {\pi -4\,\sqrt {\pi +4\,{|\B|}^{2}}|\B|+8\,{|\B|}^{2}}{\pi }\right)}
\end{equation}
For very small values of $|\B|$, the correction to the power spectrum varies linearly with $|\B|$ for such excited states
\begin{equation}\label{pwr-lin}
P_S\simeq\frac{H^2}{4 \pi^2\epsilon} (1-\frac{4}{\sqrt{\pi}}|\B|)
\end{equation}

\section{DBI Inflation}

The perturbations in the DBI inflation \cite{Silverstein:2003hf,Alishahiha:2004eh} satisfy the following equation of motion:
\begin{equation}\label{DBI-pert}
v''_{k}+c_s^2 k^2 v_k-\frac{z''}{z} v_k=0
\end{equation}
where $v_k$ is related to the Fourier transform of the gauge-invariant perturbations of the comoving hypersurface, $u_k$, through the following relation
\begin{equation}\label{u-v}
v_k\equiv z \zeta_k,~~~~~~~~~~~~z\equiv\frac{a\sqrt{2\epsilon}}{c_s}
\end{equation}
Above, $c_s$ is the speed of sound  and $\epsilon \equiv - \dot{H}/H^2$ is the first generalized slow-roll parameter. In de-Sitter space, $a=-\frac{1}{H\eta}$, the solution to differential equation \eqref{DBI-pert} could be expressed in terms of Hankel functions of the first and second kind:
\begin{equation}\label{DBI-sol}
v_k(\eta)=\A\sqrt{-\eta} H_{3/2}^{(1)}(-k c_s \eta)+\B \sqrt{-\eta} H_{3/2}^{(2)}(-k c_s \eta)
\end{equation}
The Wronskian condition for the general solution of \eqref{DBI-sol} reduces to the previously derived equation, eq.\eqref{Wronskian}. The two-point function for the general solution \eqref{DBI-sol} is
\begin{equation}\label{}
P_S=\frac{H^2 |\A-\B|^2}{2 \pi^3 c_s \epsilon}
\end{equation}
Requiring the primordial power spectrum be unmodified, one again obtains  equation \eqref{AB-scale-invariant}. Since the two equations that determine the calm excited states
(that leave the two-point function invariant) for DBI inflation are the same as that for slow-toll inflation, one can conclude that this class of calm excited states are common among both slow-roll and DBI inflationary scenarios.

It is also useful to  find the enhancement of flattened configurations for such calm excited states in DBI inflation. The three-point function in the limit that $\tk_j\rightarrow 0$ is calculated in the appendix. The leading terms in the limit where  one has an flattened configuration is:
\begin{equation}\label{tttot}
\left.\Delta\langle
\zeta(t,\textbf{k}_1)\zeta(t,\textbf{k}_2)\zeta(t,\textbf{k}_3)\rangle\right|_{\tk_j\rightarrow 0}=\mE_2 f_3(k_i) (k_t \eta_0)^3+\mE_3 f_2(k_i) (k_t  \eta_0)^2+\mE_2 f_1(k_i) k_t \eta_0,
\end{equation}
with the explicit expressions for $\mE_2$ and $\mE_3$ given in the appendix. The factors $f_j(k_i)$, $j=1\cdots3$, could also be obtained from the results in the appendix. For the calm excited states that leave the two-point function intact, one has:
\begin{eqnarray}\label{mE-dep-2-pt-calm}
  \mE_2 &=& \mp \frac{\pi^{\frac{3}{2}}}{4}{\rm sign}(y_1)|\B| \\
  \mE_3&=& -\frac{3}{2}\pi |\B|^2.
\end{eqnarray}
From the above dependence of $\mE_i$ on $|\B|$, one sees that even assuming that the excited states maximally violate the adiabaticity condition, {\it i.e.} when condition \eqref{sigma-bound} is saturated, the quadratic and linear order (in $k_t \eta_0$) corrections  will be negligible.
  This is because any probe of non-gaussianity through the CMB will be two-dimensional and such a projection from three dimensions to two, will cause one to lose a factor of $|k_t\eta_0|$  in the enhancement \cite{Holman:2007na}. Thus the real enhancement in this case comes from the cubic term. The enhancement is always nonzero for these states, as $\mE_2$ is proportional to $|\B|$.

\subsection{Absence of Excited States with No Enhancement for the Flattened Configurations in DBI Inflation}

One may wonder if in DBI inflation it is possible to identify excited states for which the enhancement of flattened configurations is identically zero. It is easy to verify that there are no such excited states in DBI inflation, since the factors $\mE_2$ and $\mE_3$ could not be set both to zero simultaneously. One can identify at most the excited states for which the {\it leading} contribution to the enhancement in the non-gaussianity is zero, {\it i.e.} $\mE_2=0$. Then the term linearly proportional to $k_t\eta_0$ will also be automatically zero. Nonetheless, the quadratic term proportional to  $\mE_3$ will survive. Since $\mE_2\propto C_4=0$, the excited state in DBI inflation for which the leading enhancement for the flattened configurations is zero, are the same excited states in slow-roll inflation for which there is no enhanced nongaussianity for the flattened configurations. The coefficient of the quadratic term for such states is
\begin{equation}\label{me3-3pt-calm}
\mE_3=-4|\B|^4+2|\B|^3\sqrt{4 |\B|^2+\pi}-\pi|\B|^2+\frac{\pi|\B|}{4}\sqrt{4 |\B|^2+\pi}.
\end{equation}
For small values of $|\B|$, $\mE_3$ could be expanded as:
\begin{equation}\label{me3-exp}
\mE_3\simeq \frac{\pi^{3/2}|\B|}{4}+\cdots
\end{equation}
Thus after taking into account that a two-dimensional projection of the three-point function will suppress the enhancement by a factor of $k_t\eta_0$, the observable enhancement is
\begin{equation}\label{obs-3pt-calm}
\left.\frac{\Delta\langle
\zeta(t,\textbf{k}_1)\zeta(t,\textbf{k}_2)\zeta(t,\textbf{k}_3)\rangle}{\langle
\zeta(t,\textbf{k}_1)\zeta(t,\textbf{k}_2)\zeta(t,\textbf{k}_3)\rangle}\right|_{\tk_j\rightarrow 0}\simeq |\B| k_t\eta_0
\end{equation}
The factor $k_t \eta_0$ is the ratio of physical momentum at the beginning of inflation to the Hubble scale which may be as large as $M/H$. Assuming that $|\B|$ takes its maximum value allowed by the constraint in  \eqref{sigma-bound}, the enhancement will be proportional to
\begin{equation}\label{obs-3-pt}
\left.\frac{\Delta\langle
\zeta(t,\textbf{k}_1)\zeta(t,\textbf{k}_2)\zeta(t,\textbf{k}_3)\rangle}{\langle
\zeta(t,\textbf{k}_1)\zeta(t,\textbf{k}_2)\zeta(t,\textbf{k}_3)\rangle}\right|_{\tk_j\rightarrow 0}\simeq  \sqrt{\epsilon \eta'} \frac{M_{\rm P}}{M}.
\end{equation}
Thus, the observable enhancement is only negligible if
\begin{equation}\label{M-Mp}
M\gtrsim {\rm min}\{\sqrt{\epsilon \eta'}, \sqrt{\epsilon}\} M_{\rm P}
\end{equation}
For an inflationary scenario with $\epsilon\simeq \eta'\sim 10^{-2}$, the enhancement from the flattened configuration is absent if
\begin{equation}\label{m-mp-example}
M \gtrsim 10^{-2} M_{\rm P},
\end{equation}
Otherwise, the enhancement from the flattened configuration can be considerable.

Here, we pause to compare the enhancement we found with the results of some
previous works. In \cite{Meerburg:2009fi},
the coefficient $\A$ was simply set to
$\A=\frac{\sqrt{\pi}}{2}$ throughout the calculation \footnote{It was noted in \cite{Meerburg:2009fi} that the phase of $\beta_k$ can have significant observational consequences. In our current work reported here, we found that $\alpha_k$ and its phase, consistent with and demanded by the Wronskian condition, can have even more dramatic observational effects.}
without regards to
the Wronskian condition.
We believe this leads
to an overestimate of the
the enhancement factor for the flattened configurations to be proportional to $(k_t\eta_0)^3$ (before the two-dimensional projection). Our analysis shows that, not only $\B$, but $\A$ is important in estimating the effect of excited states in the three-point function.

\section{Conclusion}

It is often expected that any deviation from the Bunch-Davies initial state of inflation would lead to a modification of the power spectrum. Here we showed that this expectation is not necessary true: we identified a two-parameter family of initial states in slow-roll and DBI inflation for which there is no modification to the two-point function. If it were not the backreaction constraint, the deviation could be arbitrarily large. All such excited states are distinguishable from their three-point functions
as they lead to an enhanced non-Gaussianity
in the flattened configurations. A correct estimation of
such enhancement requires
 one
to take into account the
Wronskian condition
which has so far been neglected in some earlier analysis.
Taking into account this condition, we identified a class of excited states in slow-roll inflation that maximally violate the adiabaticity condition
for which the enhancement for flattened configurations is identically zero. In DBI inflation all excited states would lead to an  enhancement for flattened configurations. This shows that for slow-roll inflation, a combination of two and three-point function measurements are necessary to detect the initial state. However in DBI inflation, the three-point function by itself can reveal the nature of the initial states for inflation. It would be interesting to see if such calm excited states could be motivated from a microphysical point of view, and whether additional features may arise in concrete models/scenarios. To present explicit expressions in this paper,
we considered the simple case in which the the Bogolyubov coefficients are scale-independent. More realistic situations may involve coefficients in which they are scale-dependent
but the results will be highly model dependent. Having a concrete model in hand,
however,
one can easily generalize our computations to incorporate the scale dependence of the Bogolyubov coefficients.
Also, it would be of great interest to extend our current study to higher point functions such as the trispectra especially for models
 with non-canonical kinetic terms where the non-Gaussianities generated can be observably large \cite{Chen:2009bc,Arroja:2009pd,Huang:2006eha,Arroja:2008ga}.

\newpage

\vspace{1cm}
\appendix{\large{\textbf{Appendix: Three-point Function for General Excited States in DBI Inflation}}}\label{append}
\vspace{0.4cm}

For a general single field inflationary scenario with the following Lagrangian:
\begin{equation} \label{general}
S=\frac{1}{2}\int d^4x \sqrt{-g} \left[M_{pl}^2R +
2P(X,\phi)\right]~,
\end{equation}
the interaction Hamiltonian for the curvature perturbations of the comoving hypersurface, $\zeta$, to $\mathcal{O}(\epsilon^2)$ is \cite{Chen:2006nt}:
\begin{eqnarray} \label{action4}
H_{int}(t)&=&-\int
d^3x\{ -a^3 (\Sigma(1-\frac{1}{c_s^2})+2\lambda)\frac{\dot{\zeta}^3}{H^3}+
\frac{a^3\epsilon}{c_s^4}(\epsilon-3+3c_s^2)\zeta\dot{\zeta}^2
\nonumber \\ &+&
\frac{a\epsilon}{c_s^2}(\epsilon-2s+1-c_s^2)\zeta(\partial\zeta)^2-2a\frac{\epsilon}{c_s^2}\dot{\zeta}(\partial
\zeta)(\partial \chi) \}~.
\end{eqnarray}
$\phi$ above is the inflaton field and $X=-\frac{1}{2}g^{\mu\nu}\partial_{\mu}\phi \partial_{\nu}\phi$. $\chi$ satisfies the following equation:
\begin{equation}\label{}
\partial^2 \chi= a^2 \frac{\epsilon}{c_s^2} \dot \zeta ~,
\end{equation}
Also
\begin{eqnarray}
\Sigma&=&X P_{,X}+2X^2P_{,XX}  = \frac{H^2\epsilon}{c_s^2} ~,\\
\lambda&=& X^2P_{,XX}+\frac{2}{3}X^3P_{,XXX} ~.
\label{lambda}
\end{eqnarray}
It is useful to define the ``speed of sound'' $c_s$ as
\begin{eqnarray}
c_s^2 = \frac{dP}{dE}= \frac{P_{,X}}{P_{,X}+2X P_{,XX}}
\end{eqnarray}
and some ``slow variation parameters'' as
in standard slow roll inflation
\begin{eqnarray} \label{small}
\epsilon&=& -\frac{\dot{H}}{H^2}=\frac{X P_{,X}}{\Mpl^2 H^2}~, \nonumber \\
\eta &=& \frac{\dot{\epsilon}}{\epsilon H}~, \nonumber \\
s &=& \frac{\dot{c_s}}{c_s H} ~.
\end{eqnarray}
The expectation value of the three-point function in the interaction picture is given by
\begin{eqnarray}
\langle
\zeta(t,\textbf{k}_1)\zeta(t,\textbf{k}_2)\zeta(t,\textbf{k}_3)\rangle=
-i\int_{t_0}^{t}dt^{\prime}\langle[
\zeta(t,\textbf{k}_1)\zeta(t,\textbf{k}_2)\zeta(t,\textbf{k}_3),H_{int}(t^{\prime})]\rangle ~,
\end{eqnarray}
Below, we will find the contribution of each term in the interaction Hamiltonian:

Contribution from $\dot{\zeta}^3$ term could be calculated in the following way. We denote
$k_t=k_1+k_2+k_3$, and find
\begin{eqnarray}
T_1&=& -i(c_s^2-1+\frac{2\lambda
c_s^2}{\Sigma})\frac{H^2\epsilon}{c_s^4}u(0,\textbf{k}_1)
u(0,\textbf{k}_2)u(0,\textbf{k}_3)
\int
_{\eta_0}^{0}\frac{a d\eta}{H^3} \nonumber \\ &\times&
(6\frac{d u^*(\eta,\textbf{k}_1)}{d \eta}\frac{d
u^*(\eta,\textbf{k}_2)}{d \eta}\frac{d u^*(\eta,\textbf{k}_3)}{d
\eta})
(2\pi)^3 \delta^3(\sum_i\bk_i) + {\rm c.c.}. \\
\label{IntFirst}
\end{eqnarray}
Using the following equations
\begin{eqnarray}\label{u0}
u(0,\textbf{k})&=&-\frac{i(\A-\B)}{\sqrt{\pi k^3 c_s\epsilon}}\\
\frac{d u^*(\eta,\mathbf{k}_1)}{d \eta}&=&-\frac{i H \eta \sqrt{k c_s^3}(-\OA \exp(i k c_s\eta))+\OB \exp(-i k c_s\eta))}{\sqrt{\pi \epsilon}}
\end{eqnarray}
$T_1$ will become:
\begin{eqnarray}\label{T1-second}
T_1&=&\frac{96}{k_1 k_2 k_3} (c_s^2-1+\frac{2\lambda
c_s^2}{\Sigma})\frac{H^4}{c_s \epsilon^2} \delta^3(\sum_i\bk_i) \left[\Re\left(i (\A-\B)^3(\OA^3\int_{\eta_0}^0 \eta^2 \exp(i k_t c_s\eta)-\OB^3 \int_{\eta_0}^0 \eta^2 \exp(-i k_t c_s\eta))\right)\right.\nonumber \\
&+&\left.\sum_{j=1}^{3}\Re\left(i (\A-\B)^3(-\OA^2 \OB \int_{\eta_0}^0 \eta^2 \exp(i \tilde{k}_j c_s\eta)+\OB^2 \OA \int_{\eta_0}^0 \eta^2 \exp(-i \tilde{k}_j c_s\eta))\right)\right]+\mathrm{perm.}
\end{eqnarray}
Upon integration yields:
\begin{eqnarray}\label{T1-third}
T_1&=&\frac{\pi H^4 (c_s^2-1+\frac{2\lambda
c_s^2}{\Sigma})}{c_s^4 \epsilon^2 k_1 k_2 k_3}\delta^3(\sum_i\bk_i) \left[\frac{1}{k_t^3}\left(-24\mE_1+\cos(\xi_t)(24 \mE_1-72\xi_t\mE_2-12 \mE_1 \xi_t^2)+ \sin(\xi_t) (72 \mE_2+24\mE_1 \xi_t-36 \mE_2 \xi_t^2)\right)\right.\nonumber\\
&+&\left. \sum_{j=1}^{3}\frac{1}{\tilde{k}_j^3}\left(24\mE_3-\cos(\xi_j)(-24\mE_3+24 \mE_2\xi_j+12 \mE_3\xi_j^2)+\sin(\xi_j)(-24\mE_2-24\mE_2\xi_j+ 12 \mE_3\xi_j^2)\right)\right]
\end{eqnarray}
where $\xi_t\equiv k_t c_s\eta_0$, $\xi_j\equiv \tk_j c_s\eta_0$ and $\mE_i$, $i=1..3$ are as follows:
\beqa
\mE_1&=& 2 \A^2 \OA^2 - 3 \A \OA^2 \B + 3 \OA^2 \B^2 - 3 \A^2 \OA \OB +2 \A \OA \B \OB - 3 \OA \B^2 \OB + 3 \A^2 \OB^2 - 3 \A \B \OB^2 +2 \B^2 \OB^2\nonumber\\
\mE_2&=&i (\A - \B) (\OA - \OB) (-\OA \B + \A \OB)\nonumber\\
\mE_3&=&\A \OA^2 \B + \OA^2 \B^2 + \A^2 \OA \OB - 6 \A \OA \B \OB +\OA \B^2 \OB + \A^2 \OB^2 + \A \B \OB^2
\eeqa
In the limit where one of the $\tk_j$'s approaches zero, $T_1$ would behave like:
\begin{equation}\label{tT1}
\tilde{T}_1\equiv\lim_{\tk_j\rightarrow 0} T_1=\frac{4\pi \mE_2 H^4\left(\frac{2\lambda}{\Sigma}-1+c_s^2\right)}{c_s^4 \epsilon^2 k_1 k_2 k_3}c_s^3 \eta_0^3
\end{equation}

For the contribution of $\zeta\dot{\zeta}^2$ term, one finds
\begin{eqnarray}
T_2&=&i\frac{\epsilon}{c_s^4}(\epsilon-3+3c_s^2)u(0,\textbf{k}_1)
u(0,\textbf{k}_2)u(0,\textbf{k}_3) \int
_{\eta_0}^{0} a^2 d\eta \nonumber \\ &\times&
2(u^*(\eta,\textbf{k}_1)\frac{d u^*(\eta,\textbf{k}_2)}{d
\eta}\frac{d u^*(\eta,\textbf{k}_3)}{d \eta}+\textrm{sym})
(2\pi)^3 \delta^3(\sum_i\bk_i) +{\rm c.c.}
\label{IntSecond}
\end{eqnarray}
Plugging the expressions for $u(0,\textbf{k})$ and $\frac{d u^*(\eta,\textbf{k})}{d
\eta}$ into the integral yields
\beqa
T_2&=&\frac{32 H^4 (\epsilon-3+3c_s^2)}{c_s^3 \epsilon^2 k_1^3 k_2 k_3}\Re\left[(\A-\B)^3 \int_{\eta_0}^{0}\left(\OA^3(k_1 c_s \eta+i)e^{ik_t c_s\eta}+\OB^3(k_1 c_s \eta-i) e^{-ik_t c_s\eta}\right.\right.\nonumber\\
&&+\OA^2\OB (k_1 c_s \eta-i) e^{i\tk_1 c_s \eta}+\OA\OB^2(k_1 c_s\eta+i) e^{-i \tk_1 c_s \eta}\nonumber\\
&&+\left.\left.\sum_{j=2}^3 \left(-\OA^2 \OB(k_1 c_s \eta+i)e^{i \tk_j c_s \eta}-\OA\OB^2(k_1 c_s \eta-i)e^{-i\tk_j c_s \eta}\right)\right)\right]+\mathrm{perm.}
\eeqa
Performing the integration, one finds:
\beqa
T_2&=&\frac{\pi H^4 (\epsilon-3+3c_s^2)}{c_s^2 \epsilon^2 k_1 k_2 k_3}\times\nonumber\\
&& \left[\left(4\mE_1\left(\frac{1}{l_1 k_t^2}+\frac{1}{k_t l_2^2}\right)+\cos(\xi_t)\left(-4\mE_1\left(\frac{1}{l_1 k_t^2}+\frac{1}{k_t l_2^2}\right)+12\mE_2 \xi_t \frac{1}{l_1 k_t^2}\right)+\sin(\xi_t)\left(-12\mE_2\left(\frac{1}{k_t^2 l_1}+\frac{1}{k_t l_2^2}\right)-4\frac{\xi_t\mE_1}{l_1 k_t^2}\right)\right)\right. \nonumber\\
&+&\left(4\mE_3\frac{k_1-\tk_1}{k_1^2 \tk_1^2}+\cos(\xi_1)\left(-4\mE_3\frac{(k_1-\tk_1)}{k_1^2 \tk_1^2}+\frac{4 \mE_2 \xi_1}{k_1\tk_1^2} \right)+\sin(\xi_1)\left(-4\mE_2\frac{k_1-\tk_1}{\tk_1^2 k_1^2}-\frac{4\mE_3 \xi_1}{k_1 \tk_1^2}\right)\right.\nonumber\\
&+&\left.\left.\sum_{j=2}^{3} \left(-4\mE_3\frac{k_1+\tk_j}{k_1^2 \tk_j^2}+\cos(\xi_j)\left(4\mE_3\frac{(k_1+\tk_j)}{k_1^2 \tk_j^2}-\frac{4 \mE_2 \xi_j}{k_1\tk_j^2} \right)+\sin(\xi_j)\left(4\mE_2\frac{k_1+\tk_j}{\tk_j^2 k_1^2}+\frac{4\mE_3 \xi_j}{k_1 \tk_j^2}\right)\right)+{\rm perm.}\right)\right].
\eeqa
where
\begin{eqnarray}
  \frac{1}{l_1} &\equiv& \sum_{i=1}^3 \frac{1}{k_i}\\
\frac{1}{l_2^2} &\equiv& \sum_{i=1}^3 \frac{1}{k_i^2}\\
\end{eqnarray}
In the limit $\tk_j\rightarrow 0$, we have:
\begin{equation}\label{tt2}
\tilde{T}_2\equiv\lim_{\tk_j\rightarrow 0} T_2= \frac{2\pi (-3+3 c_s^2+\epsilon)}{c_s^4 \epsilon^2 k_1 k_2 k_3}\left[\frac{2\mE_2}{l_2^2}c_s\eta_0+
\mE_3\left(\frac{1}{k_j}-\frac{1}{k_{j+1}}-\frac{1}{k_{j+2}}\right)c_s^2 \eta_0^2\right]
\end{equation}

For the $\zeta(\partial\zeta)^2$ term, one obtains:
\begin{eqnarray}
T_3&=&\frac{2i \epsilon}{c_s^2}(\epsilon-2s+1-c_s^2)u(0,\textbf{k}_1)
u(0,\textbf{k}_2)u(0,\textbf{k}_3) \int
_{\eta_0}^{0} a^2 d\eta \nonumber \\ &\times&
( (\textbf{k}_2.\textbf{k}_3) u^*(\eta,\textbf{k}_1)u^*(\eta,\textbf{k}_2) u^*(\eta,\textbf{k}_3)+\textrm{sym})
(2\pi)^3 \delta^3(\sum_i\bk_i) +{\rm c.c.}
\label{IntThird}
\end{eqnarray}
Plugging the expressions for the relevant parameters into the above, one obtains:
\beqa
T_3&=&-\frac{32 H^4 (\epsilon-2s+1-c_s^2)(\textbf{k}_2.\textbf{k}_3)}{c_s^5 e^2 (k_1 k_2 k_3)^3} \Re\left[(\A-\B)^3 \left(\int_{\eta_0}^0 \frac{1}{\eta^2} \left(\OA^3 (k_1 c_s\eta+i)(k_2 c_s \eta+i)(k_3 c_s \eta+i)e^{ik_t c_s \eta}\right.\right.\right.\nonumber\\
&&+
\left.\left.\left.\OB^3 (k_1 c_s\eta-i)(k_2 c_s \eta-i)(k_3 c_s \eta-i)e^{-ik_t c_s \eta}\right.\right.\right.\nonumber \\
&&+
\left.\left.\left.
\left(\OA^2\OB (k_1 c_s\eta+i)(k_2 c_s \eta+i)(k_3 c_s \eta-i) e^{i\tk_j c_s\eta}+\OA\OB^2(k_1 c_s\eta-i)(k_2 c_s \eta-i)(k_3 c_s \eta+i) e^{i\tk_j c_s\eta}+{\rm perm.}\right)
\right)\right)\right]\nonumber\\
&& +{\rm perm.}
\eeqa
Upon integration, one finds:
\beqa
T_3&=&\frac{\pi H^4 (-1+c_s^2-\epsilon+2s)}{c_s^4 \epsilon^2}\left(\frac{2(k_1^2+k_2^2+k_3^2)}{(k_1 k_2 k_3)^3 c_s} \left[\left(\left.\frac{-3\mE_2 \cos(\xi_t)+\mE_1 \sin(\xi_t)}{\eta}- \left( \frac{-\mE_2 \cos(\xi_j)+\mE_3 \sin(\xi_j)}{\eta}+{\rm perm.}\right)\right)\right|_{\eta_0}^0\right]\right.\nonumber\\
&&\left.-\frac{(k_1^2+k_2^2+k_3^2)}{(k_1 k_2 k_3)^2}\left(\frac{1}{k_t l_1}+\frac{1}{k_t^2}\right)(-3 \mE_2 \sin(\xi_t)+\mE_1(1-\cos(\xi_t))\right.\left.-\frac{2}{ (k_1 k_2 k_3)^2 k_t^2}\left(\mE_1 \sin(\xi_t)+3 \mE_2(1-\cos(\xi_t))\right)\xi_t \right.\nonumber\\
&&+2(k_1^2+k_2^2+k_3^2)\left(
\frac{\tk_3^2(k_1+k_2)+\tk_3(k_1^2+k_2^2)-k_1 k_2 (k_1+k_2)}{(k_1 k_2 k_3)^3 \tk_3^2} \left(\mE_3(1-\cos(\xi_3))-\mE_2 \sin(\xi_3)\right)+{\rm perm.}\right)\\
&&\left. +2 (k_1^2+k_2^2+k_3^2)\left(\left(\frac{1}{(k_1 k_2 k_3)^2 \tk_3^2}\left(-\mE_2\cos(\xi_3)+\mE_3 \sin(\xi_3)\right)\right)\xi_3+{\rm perm.}\right)\right)
\eeqa
The contribution of the term in the first line is divergent when the upper limit of integral goes to zero. To make sense of it one should consider that the actual observation is made at the time of last scattering which interchanges the upper limit of integration with the corresponding conformal time at the last scattering surface, $\eta^{\ast}$. In the limit of $\tk_j\rightarrow 0$, we have:
\begin{equation}\label{}
\tilde{T}_3\equiv\lim_{\tk_j\rightarrow 0} T_3=\frac{2\pi H^4 (-1+c_s^2-\epsilon+2s)(\mathbf{k}_2.\mathbf{k}_3)}{(k_1 k_2 k_3)^3 c_s^4 \epsilon^2}\left[-\mE_3 k_1 k_2 k_3 c_s^2\eta_0^2+2\mE_2 (k_3^2+k_2^2-k_2 k_3)c_s\eta_0\right]+{\rm perm.}
\end{equation}

Finally the contribution of $\dot{\zeta}(\partial\zeta)(\partial \chi)$ will lead to the following integral
\begin{eqnarray}
T_4&=&-\frac{2i \epsilon^2}{c_s^4}u(0,\textbf{k}_1)
u(0,\textbf{k}_2)u(0,\textbf{k}_3) \int
_{\eta_0}^{0} a^2 d\eta \nonumber \\ &\times&
( 2(\textbf{k}_2.\textbf{k}_3) \frac{du^*(\eta,\textbf{k}_1)}{d\eta} u^*(\eta,\textbf{k}_2)\frac{du^*(\eta,\textbf{k}_3)}{d\eta}+\textrm{perm.})
(2\pi)^3 \delta^3(\sum_i\bk_i) +{\rm c.c.}
\label{IntFourth}
\end{eqnarray}
Plugging the expressions for the relevant parameters into the above integral, one obtains:
\beqa
T_4&=&-\frac{32 H^4 (\textbf{k}_2.\textbf{k}_3)}{c_s^3 e k_1 (k_2 k_3)^3} \Re\left[(\A-\B)^3 \left(\int_{\eta_0}^0 \left(\OA^3 (k_2 c_s \eta+i)e^{ik_t c_s \eta}+\OB^3 (k_2 c_s \eta-i)e^{-ik_t c_s \eta}\right.\right.\right. \\
&&+
\left.\left.\left.
\sum_{j=\{1,3\}}\left(-\OA^2\OB (k_2 c_s \eta+i) e^{i\tk_j c_s\eta}-\OA\OB^2 (k_2 c_s \eta-i) e^{i\tk_j c_s\eta}\right)\right.
\right.\right.\nonumber\\
&&+\left.\left.\left.\OA^2\OB (k_2 c_s\eta-i)e^{i\tk_2 c_s\eta}+\OA\OB^2 (k_2 c_s\eta+i)e ^{-ik_2 c_s\eta}\right)\right)\right]\nonumber\\
&&+{\rm perm.}
\eeqa
Integrating yields
\beqa
T_4&=&\frac{8\pi H^4 (\textbf{k}_2.\textbf{k}_3)}{c_s^4 \epsilon k_1 k_2^2 k_3^3}\left(\frac{1}{k_t^2}\left[-\frac{\mE_1 (k_2+k_t)}{k_2}+\cos(\xi_t)\left(\frac{\mE_1 (k_2+k_t)}{k_2}-3 \mE_2 \xi_t \right)+\sin(\xi_t)\left(\frac{3\mE_2 (k_2+k_t)}{k_2} +\mE_1 \xi_t\right)\right]\right.\nonumber\\
&&+\sum_{j=\{1,3\}}\frac{1 }{\tk_j^2}\left[-\frac{\mE_3 (k_2+\tk_j)}{k_2}+\cos(\xi_j)\left(-\frac{\mE_3 (k_2+\tk_j)}{k_2}+ \mE_2 \xi_j \right)+\sin(\xi_j)\left(\frac{-\mE_2 (k_2+\tk_j)}{k_2} -\mE_3 \xi_j\right)\right]\nonumber\\
&&+\left.\frac{1}{\tk_2^2} \left[-\frac{\mE_3 (k_2-\tk_2)}{k_2}+\cos(\xi_2)\left(\frac{\mE_3 (k_2-\tk_2)}{k_2}- \mE_2 \xi_2 \right)+\sin(\xi_2)\left(\frac{\mE_2 (k_2-\tk_2)}{k_2} -\mE_3 \xi_2\right)\right]\right)\nonumber\\
&&+{\rm perm.}
\eeqa
In the limit where $\tk_j\rightarrow 0$, we have:
\begin{equation}\label{tt4}
\tilde{T}_4\equiv \lim_{\tk_j\rightarrow 0} T_4=-\frac{4\pi H^4}{c_s^4 \epsilon k_1 k_2 k_3} \left[\mE_3\left(\frac{\textbf{k}_{j}.\textbf{k}_{j+2}}{k_{j} k_{j+2}}+\frac{\textbf{k}_{j}.\textbf{k}_{j+1}}{k_{j} k_{j+1}}-\frac{\textbf{k}_{j+1}.\textbf{k}_{j+2}}{k_{j+1} k_{j+2}}\right)c_s^2 \eta_0^2-2\mE_2 \left(\frac{\textbf{k}_{j}.\textbf{k}_{j+2}}{k_{j}^2 k_{j+2}^2}+\frac{\textbf{k}_{j}.\textbf{k}_{j+1}}{k_{j}^2 k_{j+1}^2}+\frac{\textbf{k}_{j+1}.\textbf{k}_{j+2}}{k_{j+1}^2 k_{j+2}^2}\right)c_s \eta_0 \right]
\end{equation}

\section*{Acknowledgments}

We are thankful to Xingang Chen,
Ulf Danielsson, Robert Mann, M. Sheikh-Jabbari, and Jan Pieter van der Schaar for helpful discussions. A.A. was supported by the G\"{o}ran Gustafsson Foundation. GS was supported in part by the US Department of Energy grant DE-FG-02-95ER40896, a Cottrell Scholar Award from Research Corporation, and a Vilas Associate Award.


\begin{thebibliography}{99}

\bibitem{Komatsu:2009kd}
  E.~Komatsu {\it et al.},
  arXiv:0902.4759 [astro-ph.CO].


\bibitem{Maldacena:2002vr}
  J.~M.~Maldacena,
  JHEP {\bf 0305}, 013 (2003).
  [astro-ph/0210603].

\bibitem{Acquaviva:2002ud}
  V.~Acquaviva, N.~Bartolo, S.~Matarrese {\it et al.},
  Nucl.\ Phys.\  {\bf B667}, 119-148 (2003).
  [astro-ph/0209156].

\bibitem{Chen:2006nt}
  X.~Chen, M.~-x.~Huang, S.~Kachru {\it et al.},
  JCAP {\bf 0701}, 002 (2007).
  [hep-th/0605045].

\bibitem{Cheung:2007st}
  C.~Cheung, P.~Creminelli, A.~L.~Fitzpatrick, J.~Kaplan and L.~Senatore,
  JHEP {\bf 0803}, 014 (2008)
  [arXiv:0709.0293 [hep-th]].



\bibitem{Holman:2007na}
  R.~Holman and A.~J.~Tolley,
  JCAP {\bf 0805}, 001 (2008)
  [arXiv:0710.1302 [hep-th]].

\bibitem{Okamoto:2003wk}
  T.~Okamoto, E.~A.~Lim,
  Phys.\ Rev.\  {\bf D69}, 083519 (2004).
  [astro-ph/0312284].

\bibitem{Easther:2004vq}
  R.~Easther, W.~HKinney, H.~Peiris,
  JCAP {\bf 0505}, 009 (2005).
  [astro-ph/0412613].


\bibitem{Meerburg:2009ys}
  P.~D.~Meerburg, J.~P.~van der Schaar, P.~S.~Corasaniti,
  JCAP {\bf 0905}, 018 (2009).
  [arXiv:0901.4044 [hep-th]].


\bibitem{Meerburg:2009fi}
  P.~D.~Meerburg, J.~P.~van der Schaar, M.~G.~Jackson,
  JCAP {\bf 1002}, 001 (2010).
  [arXiv:0910.4986 [hep-th]].


\bibitem{Chen:2006xjb}
  X.~Chen, R.~Easther, E.~A.~Lim,
  JCAP {\bf 0706}, 023 (2007).
  [astro-ph/0611645].

\bibitem{Chen:2008wn}
  X.~Chen, R.~Easther and E.~A.~Lim,
  JCAP {\bf 0804}, 010 (2008)
  [arXiv:0801.3295 [astro-ph]].

\bibitem{Chen:2010bka}
  X.~Chen,
  arXiv:1008.2485 [hep-th].


\bibitem{Lyth:2002my}
  D.~H.~Lyth, C.~Ungarelli and D.~Wands,
  Phys.\ Rev.\  D {\bf 67}, 023503 (2003)
  [arXiv:astro-ph/0208055].


\bibitem{Vernizzi:2006ve}
  F.~Vernizzi and D.~Wands,
  JCAP {\bf 0605}, 019 (2006)
  [arXiv:astro-ph/0603799].

\bibitem{Huang:2007hh}
  M.~x.~Huang, G.~Shiu and B.~Underwood,
  Phys.\ Rev.\  D {\bf 77}, 023511 (2008)
  [arXiv:0709.3299 [hep-th]].

\bibitem{Langlois:2008qf}
  D.~Langlois, S.~Renaux-Petel, D.~A.~Steer and T.~Tanaka,
  Phys.\ Rev.\  D {\bf 78}, 063523 (2008)
  [arXiv:0806.0336 [hep-th]].

\bibitem{Arroja:2008yy}
  F.~Arroja, S.~Mizuno and K.~Koyama,
  JCAP {\bf 0808}, 015 (2008)
  [arXiv:0806.0619 [astro-ph]].

\bibitem{Chen:2008ada}
  H.~Y.~Chen, J.~O.~Gong and G.~Shiu,
  JHEP {\bf 0809}, 011 (2008)
  [arXiv:0807.1927 [hep-th]].


\bibitem{Byrnes:2008zy}
  C.~T.~Byrnes, K.~Y.~Choi and L.~M.~H.~Hall,
  JCAP {\bf 0902}, 017 (2009)
  [arXiv:0812.0807 [astro-ph]].

\bibitem{Chen:2010qz}
  H.~Y.~Chen, J.~O.~Gong, K.~Koyama and G.~Tasinato,
  arXiv:1007.2068 [hep-th].

\bibitem{Ashoorioon:2008qr}
  A.~Ashoorioon, A.~Krause, K.~Turzynski,
  JCAP {\bf 0902}, 014 (2009).
  [arXiv:0810.4660 [hep-th]].



\bibitem{Naruko:2008sq}
  A.~Naruko and M.~Sasaki,
  Prog.\ Theor.\ Phys.\  {\bf 121}, 193 (2009)
  [arXiv:0807.0180 [astro-ph]].

\bibitem{Li:2009sp}
  M.~Li and Y.~Wang,
  JCAP {\bf 0907}, 033 (2009)
  [arXiv:0903.2123 [hep-th]].

\bibitem{Moss:2007cv}
  I.~G.~Moss and C.~Xiong,
  JCAP {\bf 0704}, 007 (2007)
  [arXiv:astro-ph/0701302].

\bibitem{Chen:2010xka}
  X.~Chen,
  Adv.\ Astron.\  {\bf 2010}, 638979 (2010)
  [arXiv:1002.1416 [astro-ph.CO]].


\bibitem{Chen:2007gd}
  B.~Chen, Y.~Wang and W.~Xue,
  JCAP {\bf 0805}, 014 (2008)
  [arXiv:0712.2345 [hep-th]].


\bibitem{Brandenberger:1999sw}
  R.~H.~Brandenberger,
[hep-ph/9910410].

\bibitem{Martin:2000xs}
  J.~Martin and R.~H.~Brandenberger,
  Phys.\ Rev.\  D {\bf 63}, 123501 (2001)
  [arXiv:hep-th/0005209].

  R.~H.~Brandenberger, J.~Martin,
  Mod.\ Phys.\ Lett.\  {\bf A16}, 999-1006 (2001).
  [astro-ph/0005432].

  J.~Martin, R.~Brandenberger,
  Phys.\ Rev.\  {\bf D68}, 063513 (2003).
  [hep-th/0305161].


\bibitem{Kempf:2000ac}
  A.~Kempf,
  Phys.\ Rev.\  D {\bf 63}, 083514 (2001)
  [arXiv:astro-ph/0009209].

  C.~S.~Chu, B.~R.~Greene and G.~Shiu,
  Mod.\ Phys.\ Lett.\  A {\bf 16}, 2231 (2001)
  [arXiv:hep-th/0011241].

  R.~Easther, B.~R.~Greene, W.~H.~Kinney and G.~Shiu,
  Phys.\ Rev.\  D {\bf 64}, 103502 (2001)
  [arXiv:hep-th/0104102].

  R.~Easther, B.~R.~Greene, W.~H.~Kinney and G.~Shiu,
  Phys.\ Rev.\  D {\bf 67}, 063508 (2003)
  [arXiv:hep-th/0110226].

  A.~Ashoorioon, A.~Kempf and R.~B.~Mann,
  Phys.\ Rev.\  D {\bf 71}, 023503 (2005)
  [arXiv:astro-ph/0410139].

  A.~Ashoorioon and R.~B.~Mann,
  Nucl.\ Phys.\  B {\bf 716}, 261 (2005)
  [arXiv:gr-qc/0411056].

  A.~Ashoorioon, J.~L.~Hovdebo and R.~B.~Mann,
  Nucl.\ Phys.\  B {\bf 727}, 63 (2005)
  [arXiv:gr-qc/0504135].

\bibitem{Kaloper:2002uj}
  N.~Kaloper, M.~Kleban, A.~E.~Lawrence and S.~Shenker,
  Phys.\ Rev.\  D {\bf 66}, 123510 (2002)
  [arXiv:hep-th/0201158].


\bibitem{Shiu:2002kg}
  G.~Shiu and I.~Wasserman,
  Phys.\ Lett.\  B {\bf 536}, 1 (2002)
  [arXiv:hep-th/0203113].

\bibitem{Danielsson:2002kx}
  U.~H.~Danielsson,
  Phys.\ Rev.\  D {\bf 66}, 023511 (2002)
  [arXiv:hep-th/0203198].

  R.~Easther, B.~R.~Greene, W.~H.~Kinney and G.~Shiu,
  Phys.\ Rev.\  D {\bf 66}, 023518 (2002)
  [arXiv:hep-th/0204129].

\bibitem{BEFT}
  K.~Schalm, G.~Shiu and J.~P.~van der Schaar,
  JHEP {\bf 0404}, 076 (2004)
  [arXiv:hep-th/0401164].

  B.~R.~Greene, K.~Schalm, G.~Shiu and J.~P.~van der Schaar,
  JCAP {\bf 0502}, 001 (2005)
  [arXiv:hep-th/0411217].

  K.~Schalm, G.~Shiu and J.~P.~van der Schaar,
  AIP Conf.\ Proc.\  {\bf 743}, 362 (2005)
  [arXiv:hep-th/0412288].

  B.~Greene, K.~Schalm, J.~P.~van der Schaar and G.~Shiu,
{\it In the Proceedings of 22nd Texas Symposium on Relativistic Astrophysics at Stanford University, Stanford, California, 13-17 Dec 2004, pp
0001}
  [arXiv:astro-ph/0503458].

  M.~G.~Jackson and K.~Schalm,
  arXiv:1007.0185 [hep-th].


\bibitem{Powell:2006yg}
  B.~A.~Powell, W.~H.~Kinney,
  Phys.\ Rev.\  {\bf D76}, 063512 (2007).
  [astro-ph/0612006].
\bibitem{Sarangi:2006yy}
  S.~Sarangi, K.~Schalm, G.~Shiu and J.~P.~van der Schaar,
  JCAP {\bf 0703}, 002 (2007)
  [arXiv:hep-th/0611277].
\bibitem{Ashoorioon:2006wc}
  A.~Ashoorioon, A.~Krause,
[hep-th/0607001].


\bibitem{Boyanovsky:2006qi}
  D.~Boyanovsky, H.~J.~de Vega, N.~G.~Sanchez,
  Phys.\ Rev.\  {\bf D74}, 123006 (2006).
  [astro-ph/0607508].



\bibitem{Silverstein:2003hf}
  E.~Silverstein, D.~Tong,
  Phys.\ Rev.\  {\bf D70}, 103505 (2004).
  [hep-th/0310221].

\bibitem{Alishahiha:2004eh}
  M.~Alishahiha, E.~Silverstein, D.~Tong,
  Phys.\ Rev.\  {\bf D70}, 123505 (2004).
  [hep-th/0404084].

\bibitem{Chen:2009bc}
  X.~Chen, B.~Hu, M.~x.~Huang, G.~Shiu and Y.~Wang,
  JCAP {\bf 0908}, 008 (2009)
  [arXiv:0905.3494 [astro-ph.CO]].

\bibitem{Arroja:2009pd}
  F.~Arroja, S.~Mizuno, K.~Koyama and T.~Tanaka,
  Phys.\ Rev.\  D {\bf 80}, 043527 (2009)
  [arXiv:0905.3641 [hep-th]].


\bibitem{Huang:2006eha}
  X.~Chen, M.~x.~Huang and G.~Shiu,
  Phys.\ Rev.\  D {\bf 74}, 121301 (2006)
  [arXiv:hep-th/0610235].

\bibitem{Arroja:2008ga}
  F.~Arroja and K.~Koyama,
  Phys.\ Rev.\  D {\bf 77}, 083517 (2008)
  [arXiv:0802.1167 [hep-th]].







\end{thebibliography}
\end{document}